# Probe effects on time decays of coherent anti-Stokes Raman signals of isolated transitions

Michele Marrocco

*ENEA, via Anguillarese 301, 00167 Rome, Italy*
*michele.marrocco@enea.it*

We show that time-resolved coherent anti-Stokes Raman signals of isolated transitions deviate from the exponential decay whose time constant is nowadays used for measuring Raman linewidths. The deviation is caused by probe pulse-shape effects that can be understood in terms of the spectral filtering created by the necessary action of laser probing. Consequent corrections to recent linewidth measurements on heteronuclear molecules support the unprecedented agreement between theoretical works and experiments.

It is known that time-resolved coherent anti-Stokes Raman scattering (CARS) generates a signal $S(\tau)$ that carries information on collisional broadening in a simple manner [1]. In short, the signal of an isolated molecular transition of frequency $\Omega$ and full width at half maximum (FWHM) $\Gamma$ takes the following time dependence

$$S(\tau) \propto |\chi(\tau)|^2 \tag{1}$$

with $\chi(\tau) \propto \vartheta(\tau)e^{-(i\Omega+\Gamma/2)\tau}$ indicating the non-linear susceptibility that makes the signal $S(\tau)$ decay linearly on a logarithmic scale. Thus, the negative slope of $\mathrm{Log}[S(\tau)]$ becomes informative of the collisional Raman linewidth $\Gamma$ in a fashion that is elegant, powerful and simple at the same time. For this reason, the approach is becoming increasingly popular to measure collisional broadenings that are otherwise more problematic to obtain. Many rotational linewidths of Raman transitions in $H_2$, $N_2$, $O_2$, $CO_2$, and $C_2H_2$ under several physical conditions have been measured in this way [2-12] and the method shows great promise of applicative extension to other molecules of spectroscopic interest.

The experimental strategy is attractive. It takes advantage of Raman excitation created by means of synchronous picosecond (ps) or femtosecond (fs) pump and Stokes laser pulses followed, after a certain delay $\tau$, by a ps laser pulse that probes the Raman coherence initiated by the first two pulses [1-12]. These are broadband whereas the probe is sufficiently narrow, both in time and frequency, to guarantee the spectral and temporal resolution needed to isolate the time decay of a single molecular rotational transition.

In this work, we demonstrate that, in addition to the simple dependence on $\Gamma$, the time decays of CARS signals of isolated transitions are sensitive to the shape of the probe laser. In other terms, Eq. (1) can only be regarded as an ideal limit that is achieved under special circumstances that have never been discussed in view of the high accuracy that we expect from such an advanced spectroscopic technique.

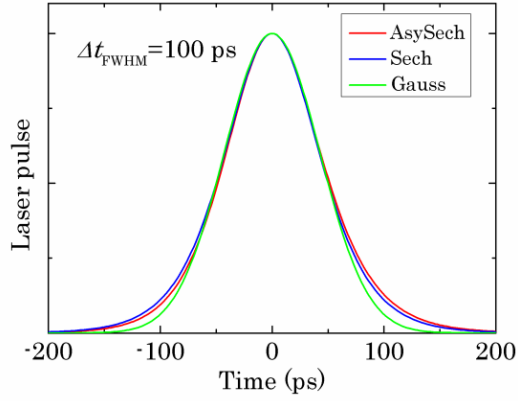

Fig. 1  Shapes of the picosecond laser pulses used in this work. The three pulses are normalized to the same maximum and the same duration (FWHM) of 100 ps.

The following analysis is based on realistic probe pulses employed in this type of measurements (Fig. 1). Their shapes vary between the extremes of the symmetric Gaussian pulse (realized by means of specific pulse shaping [4, 5]) and the asymmetric pulse of regeneratively amplified Nd:YAG lasers [13]. The latter is simulated by the square of an asymmetric hyperbolic secant (AsySech) function defined shortly. Between the two extremes, we introduce the pulse typical of passively mode-locked lasers, i.e., the square of the hyperbolic secant function (Sech) [14]. By inspection of Fig. 1, it is apparent that the pulses coincide within their central section, but they differ slightly at the rising and falling edges. Thus, the envelopes of the symmetric electric fields are $e^{-t^2/(2t_G^2)}$ for the Gauss pulse and $\text{Sech}(t/t_S)$ for the Sech pulse. The asymmetric probe-field envelope of the AsySech pulse is instead defined by $2/(e^{t/t_A}+e^{-\alpha t/t_A})$ with $\alpha > 1$ (throughout the Letter we use $\alpha = 1.2$ to simulate the very small asymmetry of Fig. 1) [15]. The choice of the time constants $t_G$, $t_S$ and $t_A$ is made in such a manner that the three pulses of Fig. 1 have the same FWHM of 100 ps.

The Fourier-transformed electric fields $\mathcal{E}_3(\omega)$ of the pulses of Fig. 1 are indispensable for calculating the time-dependent CARS signals according to a more general equation in comparison to Eq. (1) [16]

$$S_\Gamma(\Delta_{aS}-\Delta_R,\tau) \propto \left| \int_{-\infty}^{\infty} e^{-i\omega\tau} \frac{\mathcal{E}_3(\Delta_{aS}-\omega)}{\Delta_R-\omega-i\Gamma/2} d\omega \right|^2 \qquad (2)$$

and, by doing so, the time evolution of the signal takes an obvious dependence on the frequency difference between the two detunings $\Delta_{aS}$ and $\Delta_R$. The first is defined as $\Delta_{aS} = \omega_{aS}-\omega_{aS}^0$ and represents the detuning from the anti-Stokes frequency $\omega_{aS}^0 = \omega_1^0 - \omega_2^0 + \omega_3^0$ given by the CARS mixing of the carrier laser frequencies $\omega_1^0$ (pump), $\omega_2^0$ (Stokes) and $\omega_3^0$ (probe). The second is $\Delta_R = \Omega - (\omega_1^0-\omega_2^0)$ and quantifies the ordinary Raman detuning. In Eq. (2), we have made the tacit assumption of broadband Raman excitation so that we can neglect, for simplicity, the shapes of the pump and Stokes fields.

Besides, we have disregarded the interference caused by the non-resonant CARS background. Indeed, we are interested in probe delays that are no longer affected by such a spurious signal (i.e., $\tau > 200$ ps for the chosen pulses).

The result of Eq. (2) shows clearly in which way the probe shape alters the CARS signal $S_\Gamma(\Delta_{aS} - \Delta_R, \tau)$. This is understood in terms of the Fourier component of the frequency-domain non-linear susceptibility $\chi(\omega) \approx 1/(\Delta_R - \omega - i\Gamma/2)$ filtered through the spectral window supplied by the spectral profile of the probe field. In other words, the CARS decay emerges from a measurement where the field $\mathcal{E}_3(\Delta_{aS} - \omega)$ acts as if it played the role of the "instrument function" or "slit function" of a monochromator centered on $\Delta_{aS}$. This picture portrays better the idea of probing the Raman coherence that is not clearly perceived in Eq. (1).

Next, we examine the filtering effect of the probe field for the time-bandwidth limited pulses of Fig. 1 and, having established a time width of 100 ps in agreement with most of the experimental conditions [2-3, 6-12], single rotational transitions can be probed in a number of simple molecules. Considering two representative broadenings of 0.1 and 0.2 cm$^{-1}$, the on-resonance time decays $S_\Gamma(0, \tau)$ of an isolated transition are reported in Fig. 2 (higher curves refer to $\Gamma = 0.1$ cm$^{-1}$ and lower curves to $\Gamma = 0.2$ cm$^{-1}$).

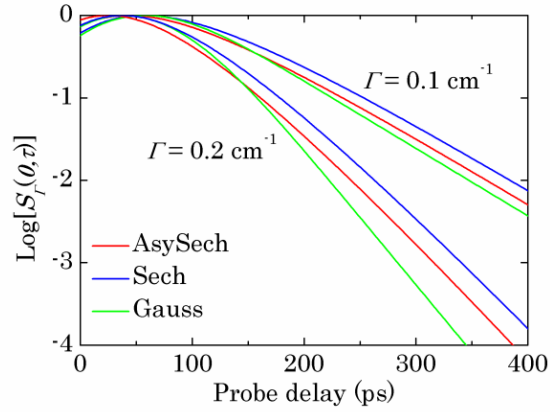

Fig. 2 Resonant CARS decays for Raman transitions with $\Gamma = 0.1$ and 0.2 cm$^{-1}$. The curves are normalized to their own maximum. The linear slopes are calculated for delays longer than 200 ps. The Gauss, Sech and AsySech signals provide different slopes, hence different Raman linewidths based on the simple assumption of Eq. (1).

The result proves that the minor changes in the time shapes of Fig. 1 have major consequences on the CARS decays. These yield different slopes if calculated for time delays where, to a good approximation, the curves seem to decrease linearly (that is, for delays $\tau > 200$ ps). These negative slopes are proportional to $\Gamma_{Eq(1)}$ extracted from Eq. (1) and we can evaluate the disagreement between $\Gamma_{Eq(1)}$ and the true linewidth $\Gamma$ by introducing the relative deviation $\Delta\Gamma/\Gamma = 100(1 - \Gamma_{Eq(1)}/\Gamma)$. This parameter is well below 1 % for the Gauss curves only ($\Delta\Gamma/\Gamma = 0.1$ and 0.2 % for $\Gamma = 0.1$ and 0.2 cm$^{-1}$, respectively). In comparison, the AsySech and Sech slopes deviates much more confirming a strong influence of the probe shape (we find for the AsySech and Sech decays that $\Delta\Gamma/\Gamma$ is 5.2 and 8.0 % at $\Gamma = 0.1$ cm$^{-1}$ and increases to 16.5 and 21.6 % at $\Gamma = 0.2$ cm$^{-1}$).

The further proof of probe pulse-shape effects is suggested by the measuring procedure in itself. As a matter of fact, the measured decay $S_m(\tau)$ is the integrated signal over the frequency span of an isolated CARS lineshape [2-12],

$$S_m(\tau) = \int_{\text{isolated CARS lineshape}} S_\Gamma(\Delta_{aS} - \Delta_R, \tau)\, d\omega_{aS}. \qquad (3)$$

This integrated signal follows the pure decay $e^{-\Gamma\tau}$ of Eq. (1) when the probe spectrum is either monochromatic (i.e., CW probe laser corresponding to $\mathcal{E}_3(\Delta_{aS} - \omega)$ given by $\delta_{DD}(\Delta_{aS} - \omega)$ with $\delta_{DD}$ the Dirac Delta function) or broadband (i.e., $\mathcal{E}_3(\Delta_{aS} - \omega)$ constant over the CARS lineshape). Both probe shapes contrast against the time and spectral resolution adopted to realize the time-resolved CARS measurements of Raman linewidths. On the other hand, $S_m(\tau)$ is nearer to the ideal decay $S(\tau)$ of Eq. (1). The integration over the CARS lineshape has, indeed, the advantage of reducing some differences observed for the resonant decays shown before.

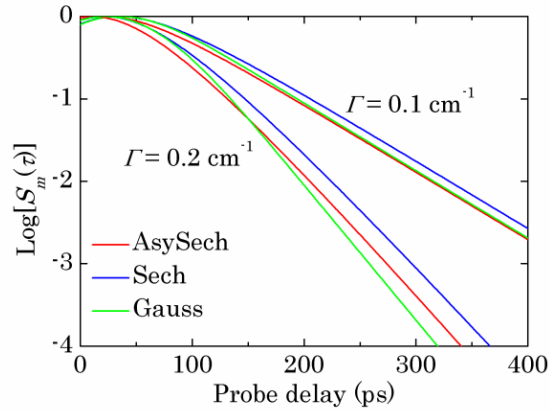

Fig. 3 Integrated CARS decays according to Eq. (3) and for Raman transitions with $\Gamma = 0.1$ and $0.2$ cm$^{-1}$. The curves are normalized to their own maximum. The linear slopes are calculated for delays longer than 200 ps. The Gauss, Sech and AsySech signals provide different slopes, hence different Raman linewidths based on the simple assumption of Eq. (1).

The advantage of employing Eq. (3) is demonstrated in Fig. 3 where $\text{Log}[S_m(\tau)]$ is plotted in complete analogy with Fig. 2. In particular, the integrated signals for $\Gamma = 0.1\,\text{cm}^{-1}$ are in better agreement with the ideal decay of Eq. (1). More exactly, the Gauss pulse is conducive to the extremely accurate value of $\Gamma$ while the relative deviation $\Delta\Gamma/\Gamma$ for the AsySech and Sech pulses reduces to 0.5 and 1.1 %, respectively. Similarly, the curves relative to $\Gamma = 0.2\,\text{cm}^{-1}$ do show smaller deviations but, if the Gauss decay approximates well the decay of Eq. (1), the other two signals are again characterized by large inaccuracies if interpreted by means of Eq. (1) (the assumed linearity results in relative deviations of 8.6 % for the AsySech pulse and 13.7 % for the Sech pulse).

The extraction of Raman linewidths from the apparent linear decays can be examined more neatly by plotting the deviation as a function of the true values of the Raman linewidth. The results are reported in Fig. 4 for on-resonance and integrated CARS signals. Here, the relative deviation $\Delta\Gamma/\Gamma$ is on a logarithmic scale in order to demonstrate the strong dependence on $\Gamma$ across the chosen range of variability and, common to all the plots, a minimum is reached around $\Gamma = 0.025$ cm$^{-1}$. But, from there, the relative deviation increases considerably at lower and higher $\Gamma$ values. Nonetheless, the Gauss probe guarantees the better agreement with Eq. (1), especially for the integrated signal of Eq. (3) that is practically indistinguishable from the ideal limit (green dashed line of Fig. 4). The other two probes show instead some critical aspects. First of all, small deviations between 0.1 and 1 % are found for integrated signals below $\Gamma = 0.1$ cm$^{-1}$. At greater $\Gamma$ values, the departure from Eq. (1) becomes progressively important up to the point where the integration of Eq. (3) ceases to be advantageous and the deviations of on-resonance and integrated signals take a similar dependence. In this limit, deviations ranging from 10 to more than 30 % are found for such probe pulses.

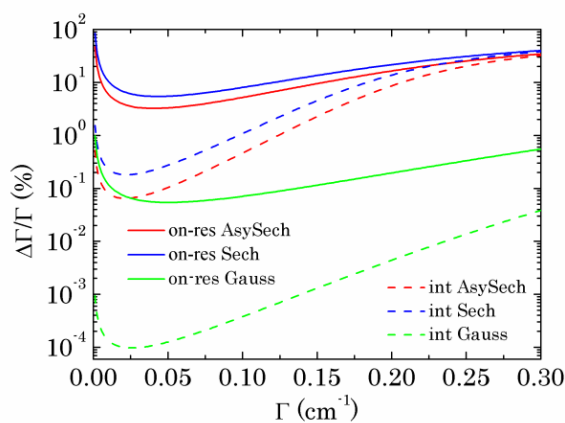

Fig. 4   Relative deviation of linewidth extraction based on Eq. (1) applied to the decays of Figs 2 (on-resonance decays, solid lines) and 3 (integrated decays, dashed lines).

The evidence of pulse-shape effects given above can be used to revise published measurements of Raman broadenings. The case of heteronuclear molecules is particularly instructive in that the reported rotational measurements of $CO_2$, and $C_2H_2$ broadenings, based on Eq. (1) [9, 10], are sensibly smaller than the values reported in previous studies on the subject [17, 18]. Since the time-resolved CARS signals were generated by means of asymmetric probe pulses of the type described by Roy et al. [13] and exemplified in Fig. 1, it is highly probable that the smaller measured linewidths may be explained by pulse-shape effects. It is then not surprising that, taking into account the deviation expected for the collisional $CO_2$ self-broadenings of Roy et al. [9], the corrected broadenings agree with the theoretical calculation of Rosenmann et al. [17]. Similar considerations can also be made for experimental $C_2H_2$ broadenings that are found smaller than earlier theoretical elaborations [10].

The discussion leading to Fig. 4 holds for the differences among the pulses of Fig. 1 and its conclusions stimulate a follow-up question on the consequences of parametric modification of the probe pulses. For example, a better spectral CARS resolution is obtained by increasing the probe pulse duration. The consequences can be viewed in the plots of Fig. 5 where probe pulses of 120 ps replace those of Fig. 1. The comparison with Fig. 4 indicates vividly that the better spectral resolution (or the longer probe duration) is detrimental to the accuracy by which the true Raman linewidth can be retrieved from linear decays in logarithmic plots. In this regard, it suffices to observe that, unlike before, the stronger probe pulse-shape effects make the integrated Gauss signal inaccurate in some measure (with relative deviation of 1 % at $\Gamma = 0.3$ cm$^{-1}$ and greater than 1 % for $\Gamma > 0.3$ cm$^{-1}$).

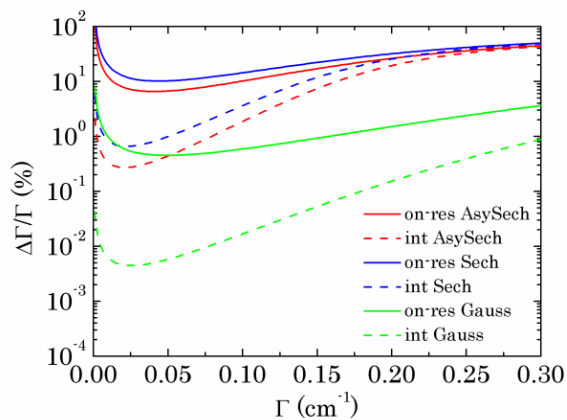

Fig. 5    Relative deviation of linewidth extraction based on Eq. (1) applied to the CARS decays generated by means of probe pulses of 120 ps duration. The vertical axis of Fig. 4 is reported here for the sake of comparison. Solid lines represent on-resonance decays and dashed lines represent integrated decays.

In conclusion, this work underlines the necessity of an evaluation of probe pulse-shape effects in time-resolved CARS measurements of Raman linewidths. These effects are generally of minor importance for Gaussian pulses but they might be worrisome when using other probe shapes with the fundamental exception of Fourier-limited probe pulses that are shaped like an exponential decay in the time domain. Although the exponential pulse has caught the attention by reason of other applications of time-resolved CARS [19, 20], the accurate theory of probe effects suggests this type of probe pulses for the full compatibility with the ideal limit of Eq. (1) [16]. Was this peculiar pulse unavailable, the option to interpret the pulse-shape dependent CARS signal would be at fingertips anyway.